\documentclass[aps,prd,preprint,nofootinbib,superscriptaddress]{revtex4-1}

\usepackage{mathtools,amsmath}
\usepackage{xcolor}
\usepackage{graphicx}
\usepackage{epsfig}
\usepackage{amssymb}
\usepackage{amsfonts}
\usepackage{slashed}
\usepackage{booktabs}
\usepackage[%
  colorlinks=true,
  urlcolor=blue,
  linkcolor=blue,
  citecolor=blue
]{hyperref}
\usepackage{siunitx}
\usepackage{multirow}
\usepackage{colortbl}
\usepackage{lipsum}  
\usepackage{supertabular}

\def\nnb{\nonumber}

 \usepackage{caption} %% For subfigure 
 \usepackage{subcaption} %% For subfigures

\begin{document}

\title{The mass and residues of radially and orbitally excited doubly heavy baryons in QCD}

\author{T.~M.~Aliev}
\email{taliev@metu.edu.tr}
\affiliation{Department of Physics, Middle East Technical University, Ankara, 06800, Turkey}

\author{S.~Bilmis}
\email{sbilmis@metu.edu.tr}
\affiliation{Department of Physics, Middle East Technical University, Ankara, 06800, Turkey}

\begin{abstract}
The mass and residues of the first radial and orbital excitations of $J=\frac{1}{2}$ baryons containing two heavy quarks (b or c) are calculated within the QCD sum rules method. In calculations, the general forms of the interpolating currents with symmetric and anti-symmetric forms with respect to the exchange of heavy quarks have been used. Our results on the spectroscopic parameters of doubly heavy baryons are compared with the predictions of other theoretical approaches.
\end{abstract}
\maketitle

\section{Introduction}
\label{sec:intro}
Quark model predicts many hadronic states containing single, doubly and triply heavy quarks. Almost all the hadronic states with single heavy quark predicted by quark model already observed in experiments. The first observation of hadrons with doubly heavy quarks ($\Xi_{cc}^{+}$) reported by SELEX Collaboration \cite{Mattson:2002vu}  in the decay $\Xi_c^{++} \rightarrow \Lambda_c^+ K^- \pi^+$ with mass ($3519 \pm 1$)~\rm{MeV}. Recently, the LHCb Collaboration \cite{Aaij:2017ueg} has announced the observation of $\Xi_{cc}^{++}$ in the $\Xi_{cc}^{++0} \rightarrow \Lambda_c^+ K^- \pi^+ \pi^+$ decay and its mass was measured as ($3624.40 \pm 0.72 \pm 0.14$)~\rm{MeV}. This discovery stimulated theoretical studies in this subject. In the framework of different methods, such as lattice QCD \cite{Padmanath:2015jea,Bali:2015lka,Brown:2014ena}, Hamiltonian Model \cite{Yoshida:2015tia}, hyper-central method \cite{Shah:2016vmd}, QCD sum rules \cite{Zhang:2008rt,Wang_2010,Aliev:2012ru,Aliev:2012iv,Bagan:1992za} the properties of doubly heavy baryons are studied . The next to leading order corrections to the perturbative part of $J^P = \frac{1}{2}^{+}$ baryon current is calculated in \cite{Wang:2017qvg}. Practically, in all studies ground state baryons masses are calculated only. In the framework of QCD sum rules~\cite{shifman1979qcd,Ioffe:1981kw}, the first radial excitations of heavy-light mesons \cite{gelhausen2014radial}, octet \cite{jiang2015radial} and decuplet baryons \cite{Aliev_2017} are studied.

In the present work, we calculate the masses and residues of radially and orbitally excited doubly heavy baryons that are estimated in the framework of the QCD sum rules method. In calculations, we used the most general form of the interpolating currents of the doubly heavy baryons with $J=\frac{1}{2}$. Note that in the quark model's notations, these states are represented by $2^2S_{1/2}$ and $1 ^2P_{1/2}$ respectively. For brevity, we denote these states as $2S$ and $1P$ correspondingly.

This work is organized as follows. In section \ref{sec:2}, the sum rules for mass and residues of excited doubly heavy baryons are derived. In section~\ref{sec:3}, we performed a numerical analysis of the sum rules for mass and residues obtained in the previous section. This section also contains the comparison of the obtained results with the ones exists in the literature.

\section{Determination of the mass and residue of the excited doubly heavy baryons with $J = \frac{1}{2}$ in QCD sum rules}
\label{sec:2}
To determine the mass, the main object in QCD sum rules is the two-point correlation function 
\begin{equation}
  \label{eq:1}
  \begin{split}
    \Pi(q^2) & = i \int d^4x~e^{iqx} \langle 0 | T \big\{ \eta(x) \bar{\eta}(0) \big\} |0 \rangle \\
             & = \Pi_1(q^2) \slashed{q} + \Pi_2(q^2),
  \end{split}
\end{equation}
where $\eta$ is the current for the relevant baryon.

The general form of the interpolating currents for the $J  = \frac{1}{2}$ doubly heavy baryons in symmetric and anti-symmetric with respect to the exchange of two heavy quarks are:
\begin{equation}
  \label{eq:2}
  \begin{split}
    \eta^S =& \frac{1}{\sqrt{2}} \epsilon^{abc} \big\{ (Q^{aT}C q^b) \gamma_5 Q{^\prime}^c + (Q{^\prime}^a C q^b) \gamma_5 Q^c \\
    +& \beta \big[ (Q{^a}^T C \gamma_5 q^b ) Q{^\prime}^c + (Q{^\prime}^c C \gamma_5 q^b) Q^c \big] \big\}, \\
    \\
    \eta^A =& \frac{1}{\sqrt{6}} \epsilon^{abc} \big\{ 2(Q^{aT} C Q{^\prime}^b) \gamma_5 q^c + (Q^a C q^b) \gamma_5 Q{^\prime}^c \\
    +& 2 \beta  (Q{^a}^T C \gamma_5 Q{^\prime}^b ) q^c + \beta (Q{^a}^T C \gamma_5 q^b) Q{^\prime}^c \\
    -& (Q{^\prime}^{aT}) C \gamma_5 q^b ) Q^c \big\}, 
  \end{split}
\end{equation}
where $a,~b,~c$ are color indices, $Q$ and $q$ are the heavy and light quarks, $C$ is the charge conjugation operator and $\beta$ is an arbitrary parameter, $T$ is the transposition. Here we would like to note that in symmetric current both heavy quarks may be identical or different while in the anti-symmetric current the two heavy quarks must be different.

According to the QCD sum rules philosophy, this correlation function is calculated in two different domain, namely in terms of hadrons, and in terms of quark and gluons in the deep Euclidean region. Then matching the coefficients of the relevant Lorentz structures and using quark-hadron duality ansatz we get the sum rules for quantity under study.

Before presenting the details of the calculations of the phenomenological part of the correlation function, we would like to make the following remark. The separation of the contributions of the negative parity baryons from the positive ones using QCD sum rules is proposed in~\cite{Jido:1996ia}. However, in this work the contribution coming from the $2S$ state is neglected and taking into account this contribution makes impossible to solve the relevant equations for determination of the mass and residues for the $2S$ and $1P$ states in the analytical form. Hence, in this study, we assumed that the interpolating current interacts simultaneously only with the ground and one of the excited ($2S$ or $1P$) states. The price we had to pay for this assumption is being not able to separate the mass of the $2S$ and $1P$ states.

Now let us turn our attention for the calculation of the phenomenological part of the correlation function. Saturating it by baryons carrying the same quantum numbers as the interpolating current and isolating the ground and $2S$ excited states we get,
\begin{equation}
  \label{eq:3}
  \Pi^{(S,A)} = \frac{\lambda^2 (\slashed{q} + m)}{- q^2 + m^2} + \frac{\lambda_1^2 (\slashed{q} + m_1)}{m_1^2 - q^2} + ...
\end{equation}
where $\lambda(\lambda_1)$, $m,(m_1)$ are the residue and the mass of ground ($2S$) state.
Note that to derive Eq.(\ref{eq:3}) we used
\begin{equation}
  \label{eq:5}
  \langle 0 | \eta | B(q,s) \rangle = \lambda u(q,s)~.
\end{equation}
The phenomenological part of the correlation function containing $1P$ state is obtained from Eq.(\ref{eq:3}) with the help of following replacements;
\begin{equation}
  \label{eq:6}
  \begin{split}
    \lambda_1 &\rightarrow \widetilde{\lambda} \\
    m_1 & \rightarrow -\widetilde{m}
  \end{split}
\end{equation}
and $\widetilde{\lambda}$ is determined as
\begin{equation}
  \label{eq:7}
  \langle 0 | \eta | \widetilde{B}(q) \rangle = \widetilde{\lambda} \gamma_5 u(q) .
\end{equation}

As we already noted that the correlation function in terms of the quarks and gluons in the deep Euclidean region, $q^2 \ll 0$, can also be calculated with the help of operator product expansion (OPE). Using the Wick theorem and performing contraction of the heavy and light quark field for symmetric and anti-symmetric currents cases, we get expressions for the correlation function from QCD side. (See also \cite{Aliev:2012ru}.)

\begin{equation}\label{eq:symmetric}
  \begin{split}
    \Pi^{S}(q^2)&= iA\epsilon_{abc}\epsilon_{a'b'c'}\int d^4x e^{i q x}\langle0\mid\Big\{-\gamma_{5}
S^{cb'}_{Q}S'^{ba'}_{q}S^{ac'}_{Q'}\gamma_{5} - \gamma_{5}S^{cb'}_{Q'}S'^{ba'}_{q}S^{ac'}_{Q}\gamma_{5} \\ 
&+ \gamma_{5}S^{cc'}_{Q'}\gamma_{5}Tr\Big[S^{ab'}_{Q}S'^{ba'}_{q}\Big] + \gamma_{5}S^{cc'}_{Q}\gamma_{5}Tr\Big[S^{ab'}_{Q'}S'^{ba'}_{q}\Big]
+ \beta\Big( -\gamma_{5}S^{cb'}_{Q}\gamma_{5}S'^{ba'}_{q}S^{ac'}_{Q'} \\ 
&- \gamma_{5}S^{cb'}_{Q'}\gamma_{5}S'^{ba'}_{q}S^{ac'}_{Q} - 
 S^{cb'}_{Q}S'^{ba'}_{q} \gamma_{5}S^{ac'}_{Q'}\gamma_{5} -S^{cb'}_{Q'}S'^{ba'}_{q}\gamma_{5} S^{ac'}_{Q}\gamma_{5} + \gamma_{5}S^{cc'}_{Q'}Tr\Big[S^{ab'}_{Q}\gamma_{5}S'^{ba'}_{q}\Big] \\ 
&+ S^{cc'}_{Q'} \gamma_{5} Tr\Big[S^{ab'}_{Q}S'^{ba'}_{q}\gamma_{5}\Big]  +
\gamma_{5}S^{cc'}_{Q}Tr\Big[S^{ab'}_{Q'}\gamma_{5}S'^{ba'}_{q}\Big] 
+S^{cc'}_{Q} \gamma_{5}Tr\Big[S^{ab'}_{Q'}S'^{ba'}_{q}\gamma_{5}\Big]\Big) \\
&+ \beta^2\Big( -S^{cb'}_{Q}\gamma_{5}S'^{ba'}_{q}\gamma_{5}S^{ac'}_{Q'} - S^{cb'}_{Q'} \gamma_{5}S'^{ba'}_{q}\gamma_{5}S^{ac'}_{Q} + S^{cc'}_{Q'}Tr\Big[S^{ba'}_{q}\gamma_{5}S'^{ab'}_{Q}\gamma_{5}\Big] \\
&+ S^{cc'}_{Q} Tr\Big[S^{ba'}_{q}\gamma_{5}S'^{ab'}_{Q'}\gamma_{5}\Big]
\Big)
\Big\}\mid 0\rangle,
  \end{split}
\end{equation}

\begin{equation}\label{eq:antisymmetric}
  \begin{split}
 \Pi^{A}(q) &=\frac{i}{6}\epsilon_{abc}\epsilon_{a'b'c'} \int d^4x e^{i q x}\langle0\mid
\Big\{2\gamma_{5}S^{cb'}_{Q}S'^{aa'}_{Q'}S^{bc'}_{q}\gamma_{5} 
+\gamma_{5}S^{cb'}_{Q}S'^{ba'}_{q}S^{ac'}_{Q'}\gamma_{5} \\
&-2\gamma_{5}S^{ca'}_{Q'}S'^{ab'}_{Q}S^{bc'}_{q}\gamma_{5} 
+\gamma_{5}S^{cb'}_{Q'}S'^{ba'}_{q}S^{ac'}_{Q}\gamma_{5}
  -2\gamma_{5}S^{ca'}_{q}S'^{ab'}_{Q}S^{bc'}_{Q'}\gamma_{5} 
+ 2\gamma_{5}S^{ca'}_{q}S'^{bb'}_{Q'}S^{ac'}_{Q}\gamma_{5} \\ 
& +4\gamma_{5}S^{cc'}_{q}\gamma_{5}Tr\Big[S^{ab'}_{Q}S'^{ba'}_{Q'}\Big]  
+\gamma_{5}S^{cc'}_{Q'}\gamma_{5}Tr\Big[S^{ab'}_{Q}S'^{ba'}_{q}\Big]
+\gamma_{5}S^{cc'}_{Q}\gamma_{5}Tr\Big[S^{ab'}_{Q'}S'^{ba'}_{q}\Big]\\
  &+\beta \Big( 2\gamma_{5}S^{cb'}_{Q}\gamma_{5}S'^{aa'}_{Q'}S^{bc'}_{q}
+\gamma_{5}S^{cb'}_{Q}\gamma_{5}S'^{ba'}_{q}S^{ac'}_{Q'} - 2\gamma_{5}S^{ca'}_{Q'}
\gamma_{5}S'^{ab'}_{Q}S^{bc'}_{q} 
+\gamma_{5}S^{cb'}_{Q'}\gamma_{5}S'^{ba'}_{q}S^{ac'}_{Q} \\
&-2\gamma_{5}S^{ca'}_{q} \gamma_{5}S'^{ab'}_{Q}S^{bc'}_{Q'}+2\gamma_{5}S^{ca'}_{q}\gamma_{5}S'^{bb'}_{Q'}
S^{ac'}_{Q} + 2S^{cb'}_{Q}S'^{aa'}_{Q'}\gamma_{5}S^{bc'}_{q}\gamma_{5}+S^{cb'}_{Q}
S'^{ba'}_{q}\gamma_{5}S^{ac'}_{Q'}\gamma_{5} \\ &-2S^{ca'}_{Q'}S'^{ab'}_{Q}\gamma_{5}
S^{bc'}_{q}\gamma_{5}
+ S^{cb'}_{Q'}S'^{ba'}_{q}\gamma_{5}S^{ac'}_{Q}\gamma_{5} - 2S^{ca'}_{q} S'^{ab'}_{Q}\gamma_{5}S^{bc'}_{Q'}\gamma_{5} +2S^{ca'}_{q}S'^{bb'}_{Q'}\gamma_{5}
S^{ac'}_{Q}\gamma_{5} \\
&+ 4\gamma_{5}S^{cc'}_{q}Tr\Big[S^{ab'}_{Q}\gamma_{5}S'^{ba'}_{Q'}\Big]
+4S^{cc'}_{q}\gamma_{5}Tr\Big[S^{ab'}_{Q}S'^{ba'}_{Q'}\gamma_{5}\Big] +\gamma_{5}S^{cc'}_{Q'}Tr\Big[S^{ab'}_{Q}\gamma_{5}S'^{ba'}_{q}\Big] \\
& + S^{cc'}_{Q'}\gamma_{5}Tr\Big[S^{ab'}_{Q}S'^{ba'}_{q}\gamma_{5}\Big] +
\gamma_{5}S^{cc'}_{Q}Tr\Big[S^{ab'}_{Q'}\gamma_{5}S'^{ba'}_{q}\Big]+S^{cc'}_{Q}
\gamma_{5}Tr\Big[S^{ab'}_{Q'}S'^{ba'}_{q}\gamma_{5}\Big]\Big)\\
& + \beta^2\Big( 2S^{cb'}_{Q}\gamma_{5}S'^{aa'}_{Q'}\gamma_{5}S^{bc'}_{q}+S^{cb'}_{Q}
\gamma_{5}S'^{ba'}_{q}\gamma_{5}S^{ac'}_{Q'}  -
2S^{ca'}_{Q'}\gamma_{5}S'^{ab'}_{Q}\gamma_{5}S^{bc'}_{q}  + S^{cb'}_{Q'}
\gamma_{5}S'^{ba'}_{q}\gamma_{5}S^{ac'}_{Q} \\
&- 2S^{ca'}_{q}\gamma_{5}S'^{ab'}_{Q}\gamma_{5}S^{bc'}_{Q'}+2S^{ca'}_{q}\gamma_{5}
S'^{bb'}_{Q'}\gamma_{5}S^{ac'}_{Q} + 
4S^{cc'}_{q}Tr\Big[S^{ba'}_{Q'}\gamma_{5}S'^{ab'}_{Q}\gamma_{5}\Big] \\
&+S^{cc'}_{Q'}
Tr\Big[S^{ba'}_{q}\gamma_{5}S'^{ab'}_{Q}\gamma_{5}\Big] 
+ S^{cc'}_{Q}Tr\Big[S^{ba'}_{q}\gamma_{5}S'^{ab'}_{Q'}\gamma_{5}\Big]
\Big)
\Big\}\mid 0\rangle. 
\end{split}
\end{equation}
where $S'=CS^TC$.  In the case of $Q\neq Q'$, the constant $A$ in the above equation takes the value $A=\frac{1}{2}$, while when $Q= Q'$  we have $A=1$ as a result of extra contractions between the
same quark fields.

As follows from Eqs.(\ref{eq:symmetric}), and (\ref{eq:antisymmetric}), in order to calculate the relevant correlation function we need the expressions of the light and heavy quark propagators. In our calculations for light and heavy quark propagators in coordinate space we used the following expressions
\begin{equation}
  \label{eq:10}
  \begin{split}
    S_q(x) =& \frac{i \slashed{x}}{2 \pi ^2 x^4} - \frac{m_q}{4 \pi^2 x^2} - \frac{\langle \bar{q}q \rangle}{12} \big( 1 - i \frac{m_q}{4}\slashed{x} \big)
    - \frac{x^2}{192}m_0^2 \langle \bar{q}q \rangle \big( 1 - i \frac{m_1 \slashed{x}}{6}\big) \\
    -& i g_s \int_0^{1} du \big[ \frac{\slashed{x}} {16 \pi^2 x^2} G_{\mu \nu}(ux) \sigma_{\mu \nu} 
    - \frac{i}{4 \pi^2 x^2} u x_\mu G_{\mu \nu}(ux) \gamma^\nu \\
    -& i \frac{m_q}{32 \pi^2} G_{\mu \nu} \sigma^{\mu \nu} \big(\ln \frac{-x^2 \Lambda^2}{4} + 2 \gamma_E \big) \big],
  \end{split}
\end{equation}

\begin{equation}
  \label{eq:11}
  \begin{split}
S_Q(x) &= {m_Q^2 \over 4 \pi^2} {K_1(m_Q\sqrt{-x^2}) \over \sqrt{-x^2}} -
i {m_Q^2 \rlap/{x} \over 4 \pi^2 x^2} K_2(m_Q\sqrt{-x^2}) \\
&- ig_s \int {d^4k \over (2\pi)^4} e^{-ikx} \int_0^1
du \Bigg[ {\rlap/k+m_Q \over 2 (m_Q^2-k^2)^2} G^{\mu\nu} (ux)
\sigma_{\mu\nu}\\
&~~\hspace{4cm}+ {u \over m_Q^2-k^2} x_\mu G^{\mu\nu} \gamma_\nu \Bigg]~,
  \end{split}
\end{equation}
where $K_i~(i=1,2)$ are the modified Bessel functions of the second order.

Having the expressions of the light and heavy quark propagators the relevant spectral densities can be calculated. The expressions of the corresponding spectral densities had already been calculated in \cite{Aliev:2012ru,Aliev:2012iv} and for completeness, we present their expressions
\begin{eqnarray}
\rho^{S}_{1}(s)&=&\frac{A}{128 \pi^4}\int_{\alpha_{min}}^{\alpha_{max}}
\int_{\beta_{min}}^{\beta_{max}}d\alpha d\beta\Bigg\{3 \mu \Bigg[\alpha\beta
\Big[5+\beta (2+5\beta)\Big]\mu+2 (-1+\alpha+\beta)(-1+\beta)^2 m_Q m_{Q'}\nnb\\
&-&6(-1+\beta^2)m_q(\beta m_Q+\alpha m_{Q'})\Bigg]\Bigg\}+
\frac{A\langle \bar q q\rangle}{16 \pi^2}\int_{\alpha_{min}}^{\alpha_{max}}d\alpha
\Bigg\{-\Bigg[(-1+\alpha)\alpha\Big[5+\beta(2+5\beta)\Big]m_q\Bigg]\nnb\\
&+&3(-1+\beta^2)\Big[(-1+\alpha)m_Q-\alpha m_{Q'}\Big]\Bigg\},
\end{eqnarray}
\begin{eqnarray}
 \rho^{A}_{1}(s)&=&\frac{1}{256 \pi^4}\int_{\alpha_{min}}^{\alpha_{max}}
\int_{\beta_{min}}^{\beta_{max}}d\alpha d\beta\Bigg\{ \mu \Bigg[3\alpha\beta
\Big[5+\beta (2+5\beta)\Big]\mu+2 (-1+\beta)\Big[(-1+\alpha+\beta)
(13\nnb\\&+&11\beta) m_Q m_{Q'}
-(1+5\beta)m_q(\beta m_Q+\alpha m_{Q'})\Big]\Bigg]\Bigg\}\nnb\\
&+&\frac{\langle \bar q q\rangle}{96 \pi^2}\int_{\alpha_{min}}^{\alpha_{max}}
d\alpha\Bigg\{-3(-1+\alpha)\alpha\Big[5+\beta(2+5\beta)\Big]m_q\nnb\\
&+&(-1+\beta)(1+5\beta)\Big[(-1+\alpha)m_Q-\alpha m_{Q'}\Big]\Bigg\},
\end{eqnarray}
\begin{eqnarray}
 \rho^{S}_{2}(s)&=&\frac{A}{128 \pi^4}\int_{\alpha_{min}}^{\alpha_{max}}
\int_{\beta_{min}}^{\beta_{max}}d\alpha d\beta\Bigg\{3 \mu 
\Bigg[3\alpha(-1+\beta^2)\mu m_{Q'}+m_Q\Big[ 3\beta(-1+\beta^2) \mu\nnb\\
&-&2\Big[5+\beta(2+5\beta)\Big]m_q m_{Q'}\Big]\Bigg]\Bigg\}+
\frac{A\langle \bar q q\rangle}{32 \pi^2}\int_{\alpha_{min}}^{\alpha_{max}}
d\alpha\Bigg\{-\Bigg[(-1+\alpha)\alpha(-1+\beta)^2\Big[3m_0^2\nnb\\
&+&4\mu'-2 s\Big]+2\Big[5+\beta(2+5\beta)\Big]m_Q m_{Q'}+6(-1+\beta^2)
m_q\Big[(-1+\alpha)m_Q-\alpha m_{Q'}\Big]\Bigg]\nnb\\&-&\frac{3}{4}m_0^2(1-\beta)^2\Bigg\},
\end{eqnarray}
\begin{eqnarray}
 \rho^{A}_{2}(s)&=&\frac{1}{256 \pi^4}\int_{\alpha_{min}}^{\alpha_{max}}
\int_{\beta_{min}}^{\beta_{max}}d\alpha d\beta\Bigg\{ \mu \Bigg[\alpha(-1+\beta)
(1+5\beta)\mu m_{Q'}+ m_{Q}\Big[\beta(-1+\beta)(1+5\beta)\mu
\nnb\\&-&6[5+\beta(2+5\beta)]m_q m_{Q'}\Big]\Bigg]\Bigg\}+\frac{\langle
\bar q q\rangle}{192 \pi^2}\int_{\alpha_{min}}^{\alpha_{max}}d\alpha
\Bigg\{-\Bigg[(-1+\alpha)\alpha(-1+\beta)(13+11\beta)\Big[3m_0^2\nnb\\
&+&4\mu'-2 s\Big]+6\Big[5+\beta(2+5\beta)\Big]m_Q m_{Q'}+2(-1+\beta)
(1+5\beta)m_q\Big[(-1+\alpha)m_Q-\alpha m_{Q'}\Big]\Bigg]\nnb\\&+&\frac{3}{2}m_0^2(1-\beta)^2\Bigg\},
\end{eqnarray}
where,
\begin{equation}
  \begin{split}
     \mu &= \frac{m_Q^2}{\alpha}+\frac{m_{Q'}^2}{\beta}-s, \\
\mu'&= \frac{m_Q^2}{\alpha}+\frac{m_{Q'}^2}{1-\alpha}-s, \\
\beta_{min}&= \frac{\alpha m_{Q'}^2}{s\alpha-m_Q^2}, \\
\beta_{max}&= 1-\alpha,\\
  \alpha_{min}&=\frac{1}{2s}\Bigg[s+m_Q^2-m_{Q'}^2  - \sqrt{(s+m_Q^2-m_{Q'}^2)^2-
4m_Q^2s}\Bigg],\\
  \alpha_{max}&= \frac{1}{2s}\Bigg[s+m_Q^2-m_{Q'}^2 + \sqrt{(s+m_Q^2-m_{Q'}^2)^2-
4m_Q^2s}\Bigg].
\end{split}
\end{equation}

The coefficients in $\slashed{q}$ or $I$ structures in QCD and phenomenological sides are matched in order to obtain the sum rules for the mass and residue of the doubly heavy baryons under consideration. In order to suppress the higher states and continuum contributions the Borel transformation over $-q^2$ is performed and finally using the quark-hadron duality ansatz, we get the following sum rules:
\begin{equation}
  \label{eq:15}
  \begin{split}
    \Pi_1^{S(A)(B)} (M^2) &= \int_{(m_Q + m_{Q^\prime})^2}^{s_0} \rho_1^{S(A)} e^{-s/M^2} ds \\
    \Pi_2^{S(A)(B)} (M^2) &= \int_{(m_Q + m_{Q^\prime})^2}^{s_0} \rho_2^{S(A)} e^{-s/M^2} ds \\
  \end{split}  
\end{equation}  
\begin{equation}
  \label{eq:12}
  \begin{split}
   \lambda^2 e^{-m^2/M^2} + \lambda_1^2 e^{-m_1^2/M^2}  &= \Pi_1^{S(A)(B)} (M^2), \\
   m \lambda^2 e^{-m^2/M^2} + m_1 \lambda_1^2 e^{-m_1^2/M^2} &=  \Pi_2^{S(A)(B)} (M^2)
  \end{split}
\end{equation}
where $\Pi_1^{(B)}$ and $\Pi_2^{(B)}$ are the Borel transformed invariant functions in the coefficients of the Lorentz structures $\slashed{q}$ and $I$ correspondingly. Eq.\ref{eq:12} contains four unknowns, namely masses of ground and excited states and their residues. In result, we need two extra equations. These equations can be obtained by taking derivatives with respect to $\frac{-1}{M^2}$ from the two equations in Eq.\ref{eq:12}. Solving these four equations for mass and residue of radially excited state we get
\begin{equation}
  \label{eq:13}
  \begin{split}
    m_1^2 = & \frac{\Pi_2^{\prime (B)} - m \Pi_1^{\prime (B)}}{\Pi_2^{(B)} - m \Pi_1^{(B)}} \\
    \lambda_1^2 =& \frac{1}{m_1^2 - m^2} \big( \Pi_1^{\prime (B)} - m^2 \Pi_1^{(B)} \big)~ e^{{m_{1}^2}/M^2}, 
  \end{split}
\end{equation}
where $\Pi_i^{\prime (B)} = \frac{d \Pi_i^{(B)}}{d(-1/M^2)}$.

Here we would like to to make the following remark. One can easily show that the residue $\lambda_1$ can also be determined from equation
\begin{equation}
  \label{eq:9}
    \lambda_1^2 = \frac{1}{m_1 - m} (\Pi_2^{(B)}  - m \Pi_1^{(B)}) e^{{m_1^2}/M^2} 
\end{equation}
The mass difference is small and for this reason determination of $\lambda_1^2$ from this expression is not reliable. Therefore, we used Eq.(\ref{eq:13}) to determine $\lambda_1^2$. However, we used Eq.(\ref{eq:9}) for determination of $\widetilde{\lambda}^2$, since for this purpose it is enough to $\lambda_1 \rightarrow \widetilde{\lambda}$, $m_1 \rightarrow - \widetilde{m}$. In result, we get,
\begin{equation}
  \label{eq:14}
    \tilde{\lambda}^2 = \frac{1}{m + \widetilde{m}} (\Pi_1^{(B)} m - \Pi_2^{(B)}) e^{{\tilde{m}^2}/M^2}. 
\end{equation}
The mass of the $1P$ state is determined from Eq.\ref{eq:13} by replacing $m_1 \rightarrow - \tilde{m}$.
To determine the mass and residue of the excited state doubly heavy baryons the mass and residue of the ground states are taken as input parameters.

\begin{table}[hbt]
  \center
  \renewcommand{\arraystretch}{1.3}
  \setlength{\tabcolsep}{8pt}
      \begin{tabular}{ccr}
        Baryon & $\sqrt{s_0}~(\rm{GeV})$ & $M^2(\rm{GeV^2})$ \\
        \midrule
        $\Xi_{cc}$ & ($4.3 \pm 0.1$) & $4 \div 7$ \\
        $\Xi_{bb}$ & ($11.1 \pm 0.1$) & $10 \div 16$ \\
        $\Xi_{cb}$ & ($7.7 \pm 0.1$) & $7 \div 11$ \\
        $\Xi_{cb}^\prime$ & ($7.9 \pm 0.1$) & $7 \div 11$ \\
        $\Omega_{cc}$ & ($4.5 \pm 0.1$) & $4 \div 7$ \\
        $\Omega_{bb}$ & ($11.2 \pm 0.1$) & $10 \div 16$ \\
        $\Omega_{cb}$ & ($7.9 \pm 0.1$) & $7 \div 11$ \\
        $\Omega_{cb}^\prime$ & ($7.9 \pm 0.1$) & $7 \div 11$ \\
      \bottomrule
      \end{tabular}
  \caption{The working domains of continuum threshold and Borel parameters $M^2$.}
  \label{tab:1}
\end{table}

\begin{table*}[hbt]
  \renewcommand{\arraystretch}{1.2}
  \setlength{\tabcolsep}{1pt}
  \begin{tabular}{lccccccccc}
    \toprule
    & Ref.\cite{Shah:2017liu} & Ref.\cite{Yoshida:2015tia} & Ref.\cite{Roberts:2007ni} & Ref.\cite{Giannuzzi:2009gh} & Ref.\cite{Valcarce:2008dr} & Ref.\cite{Ebert:2002ig} & Ref.\cite{Eakins:2012jk}  & Ref.\cite{Wang:2010it} & Our Results
    \\
    \midrule
    $\Xi_{cc}~(2S)$ & $3.920$  & $4.079$   & $4.029$  & $4.183$  & $3.976$  &$3.910$      & $4.030$  & - & $4.03 \pm 0.20$\\
    $\Xi_{cc}~(1P)$ & $3.861$  & $3.947$   & $3.910$  & $-$  & $3.880$  &$3.838$       & $4.073$ & $3.77 \pm 0.18$ & $4.03 \pm 0.20$  \\
    $\Xi_{bb}~(2S)$ & $10.609$ & $10.571$  & $10.576$ & $10.751$ & $10.482$ &$10.441$       & $10.551$ & - & $10.32 \pm 0.10$ \\
    $\Xi_{bb}~(1P)$ & $10.551$ & $10.476$  & $10.493$ & $-$& $10.406$  & $10.368$    & $10.691$ & $10.38 \pm 0.15$ & $10.32 \pm 0.10$  \\
    $\Xi_{bc}~(2S)$ & $7.263$  & $-$       & $-$      & $7.495$      & $-$      &$-$ &$7.353$ & - & $7.14 \pm 0.11$  \\
    $\Xi_{bc}~(1P)$ & $7.156$  & $-$       & $-$      & $-$      & $-$      &$-$ & $7.390$ & -  & $7.14 \pm 0.11$ \\
    $\Xi_{bc}^\prime~(2S)$ & $-$  & $-$       & $-$      & $-$ & $-$ & $-$ & $-$  & - & $7.02 \pm 0.07$ \\
    $\Xi_{bc}^\prime~(1P)$ & $-$  & $-$       & $-$      & $-$ & $-$ & $-$ & $-$  & - &$ 7.02 \pm 0.07 $  \\ 
     \bottomrule
  \end{tabular}
  \caption{The mass of $2S$ and $1P$ excitations of $\Xi_{QQ^\prime}$ doubly heavy baryons (in \rm{GeV} unit).}
  \label{tab:2}
\end{table*}

\begin{table*}[hbt]
  \centering
  \renewcommand{\arraystretch}{1.2}
  \setlength{\tabcolsep}{1pt}
  \begin{tabular}{lccccccccc}
    \toprule
    & Ref.\cite{Shah:2016vmd} & Ref.\cite{Yoshida:2015tia} & Ref.\cite{Roberts:2007ni} & Ref.\cite{Valcarce:2008dr} & Ref.\cite{Giannuzzi:2009gh} & Ref.\cite{Ebert:2002ig}  & Ref.\cite{Wang:2010it} & Our Results
    \\
    \midrule
    $\Omega_{cc}~(2S)$ & $4.041$  & $4.227$   & $4.180$  & $4.112$  & $4.268$  &$4.075$  & -     & $4.16 \pm 0.14$ \\
    $\Omega_{cc}~(1P)$ & $3.989$  & $4.086$   & $4.046$  & $-$  & $-$  &$4.002$ & $3.91 \pm 0.14$      & $4.16 \pm 0.14$   \\
    $\Omega_{bb}~(2S)$ & $10.736$ & $10.707$  & $10.693$ & $10.604$ & $10.830$ &$10.610$ & -      & $10.37 \pm 0.10$  \\
    $\Omega_{bb}~(1P)$ & $10.646$ & $10.607$  & $10.616$ & $-$  & $-$       &$10.532$ & $10.38 \pm 0.15$ & $10.37 \pm 0.10$  \\
    $\Omega_{bc}~(2S)$ & $7.480$  & $-$       & $-$      & $-$      & $7.559$      &$-$ & - & $7.20 \pm 0.11$    \\
    $\Omega_{bc}~(1P)$ & $7.386$  & $-$       & $-$      & $-$      & $-$      &$-$ & - &  $7.20 \pm 0.11$  \\
    $\Omega_{bc}^\prime~(2S)$ & $-$  & $-$       & $-$      & $-$      & $-$      &$-$ & - & $7.09 \pm 0.07$   \\
    $\Omega_{bc}^\prime~(1P)$ & $-$  & $-$       & $-$      & $-$      & $-$      &$-$ & - & $7.09 \pm 0.07$   \\
    \bottomrule
  \end{tabular}
  \caption{Same as in Table~\ref{tab:2} but for $\Omega_{QQ^\prime}$ baryons  (in \rm{GeV} unit).}
  \label{tab:3}
\end{table*}

\section{Numerical Analysis}
\label{sec:3}
In this section, we present our numerical analysis and show our results on mass and residues of doubly heavy baryons. For the  $c$ and $b$ quarks, masses have been used in the $\overline{MS}$ scheme
\begin{equation}
  \label{eq:16}
  \begin{split}
    \bar{m}_c (\bar{m}_c) =& (1.28 \pm 0.03)~\rm{GeV}   \\
    \bar{m}_b (\bar{m}_b) =& (4.16 \pm 0.03)~\rm{GeV} .
  \end{split}
\end{equation}
The values of other input parameters are:
\begin{equation}
  \label{eq:8}
  \begin{split}
    m_s(2~\rm{GeV}) =& (95^{+9}_{-3})~\rm{MeV}   \\
    m_0^2  =& (0.8 \pm 0.2)~\rm{GeV^2}   \\
    \langle \bar{q}q \rangle (1~\rm{GeV}) =& -(0.246^{+0.028}_{-0.019} )^3~\rm{GeV^3}   \\
    \langle \bar{s}s \rangle (1~\rm{GeV}) =& (0.8 \pm 0.2 )~ \langle \bar{q}q \rangle (1~\rm{GeV})   
  \end{split}
\end{equation}
The sum rules for the mass and residue for doubly heavy baryons contains three auxiliary parameters: the Borel mass $M^2$, continuum threshold $s_0$ and parameter $\beta$. Obviously mass should be independent of these parameters. For this reason, we need to find the working regions of these parameters in such a way where physical quantity exhibits independence of them. The Borel mass parameter $M^2$ and the continuum threshold $s_0$ are determined in such a way that the standard sum rule criteria, i.e. the suppression of continuum states and of higher twist contributions are satisfied. We obtained that these conditions are fulfilled in the regions presented in Table \ref{tab:1}. These values of $s_0$ includes only ground and first excited states. For ground state mass we used the results of \cite{Aliev:2012ru,Aliev:2012iv}.

\begin{table*}[tbh]
  \center
  \renewcommand{\arraystretch}{1.3}
  \setlength{\tabcolsep}{8pt}
      \begin{tabular}{cccc}
        \multirow{2}{*}{} & \multicolumn{3}{c}{Residue}  \\
        \multicolumn{1}{c}{} & $\lambda_1~(\rm{GeV^3})$ & ~~~$\tilde{\lambda}~(\rm{GeV^3}$) &$\tilde{\lambda}~(\rm{GeV^3}$) Ref.\cite{Wang:2010it} \\
        \midrule
        $\Xi_{cc}$  & $(0.144 \pm 0.064)$ & $(0.099 \pm 0.013)$  & $(0.159 \pm 0.037)$\\
        $\Xi_{bb}$  & $(0.764 \pm 0.114)$ &  $(0.576 \pm 0.046)$ & $(0.365 \pm 0.089)$\\
        $\Xi_{cb}$  & $(0.306 \pm 0.090)$ &  $(0.242 \pm 0.020)$ & - \\
        $\Xi_{cb}^\prime$ & $(0.187 \pm 0.092)$ &  $(0.176 \pm 0.084)$ & - \\
        $\Omega_{cc}$ & $(0.205 \pm 0.085)$ &  $(0.125 \pm 0.015)$ & $(0.192 \pm 0.041)$ \\
        $\Omega_{bb}$ & $(0.850 \pm 0.160)$ &  $(0.620 \pm 0.060)$ & $(0.444 \pm 0.101)$ \\
        $\Omega_{cb}$ & $(0.361 \pm 0.112)$ &  $(0.241 \pm 0.021)$ & - \\
        $\Omega_{cb}^\prime$ & $(0.255 \pm 0.125)$ &  $(0.174 \pm 0.079)$ & - \\
                \bottomrule
      \end{tabular}
  \caption{The residues of $2S$ and $1P$ excitations of doubly heavy baryons.}
  \label{tab:4}
\end{table*}

\begin{figure}[hbt]
\begin{subfigure}{\linewidth}
\centering
  \includegraphics[width=0.60\linewidth]{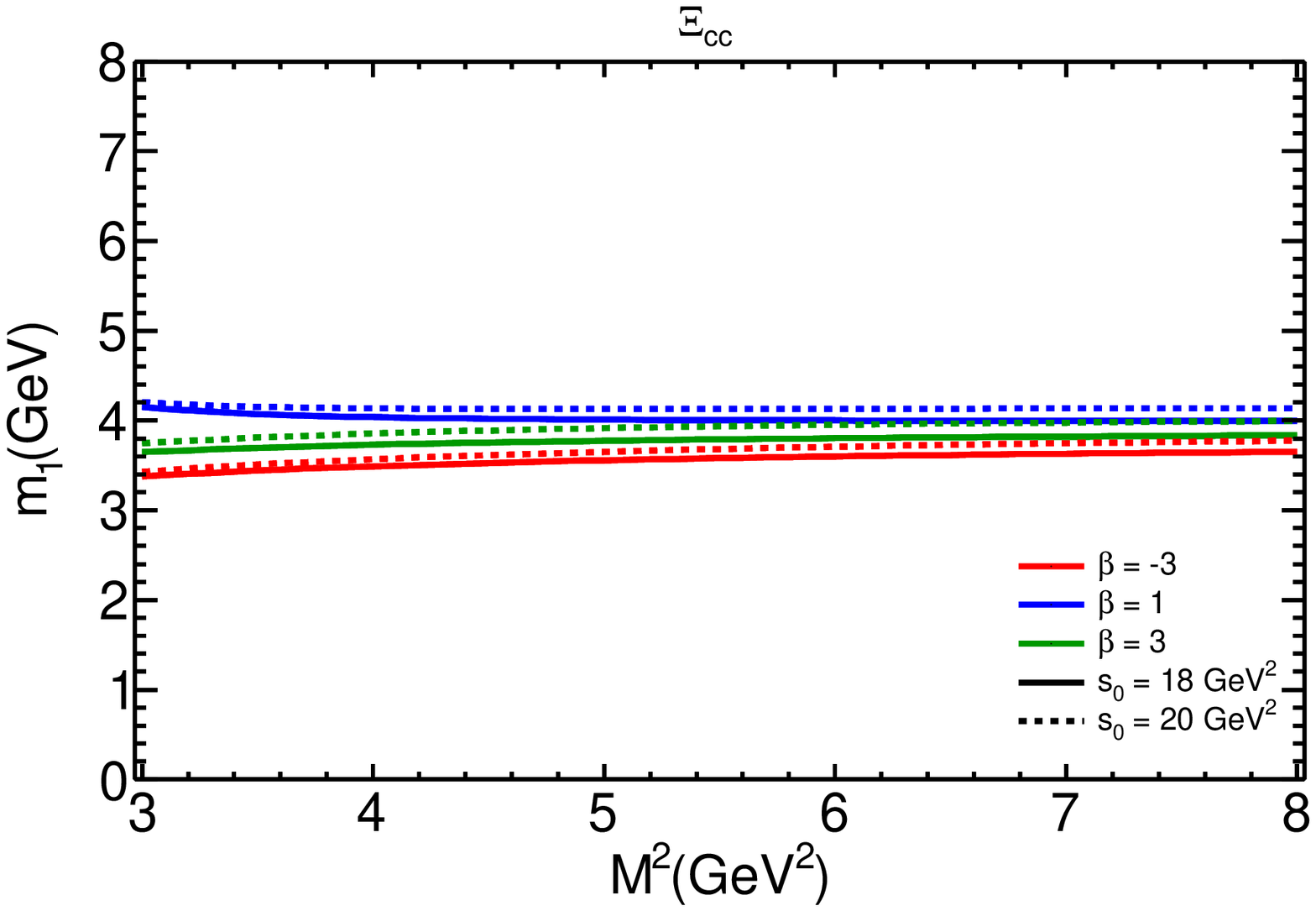}
  % \caption{~~}
\label{fig:1a}
\end{subfigure}%
 \vspace{2em}
\begin{subfigure}{\linewidth}
  \centering
  \includegraphics[width=0.60\linewidth]{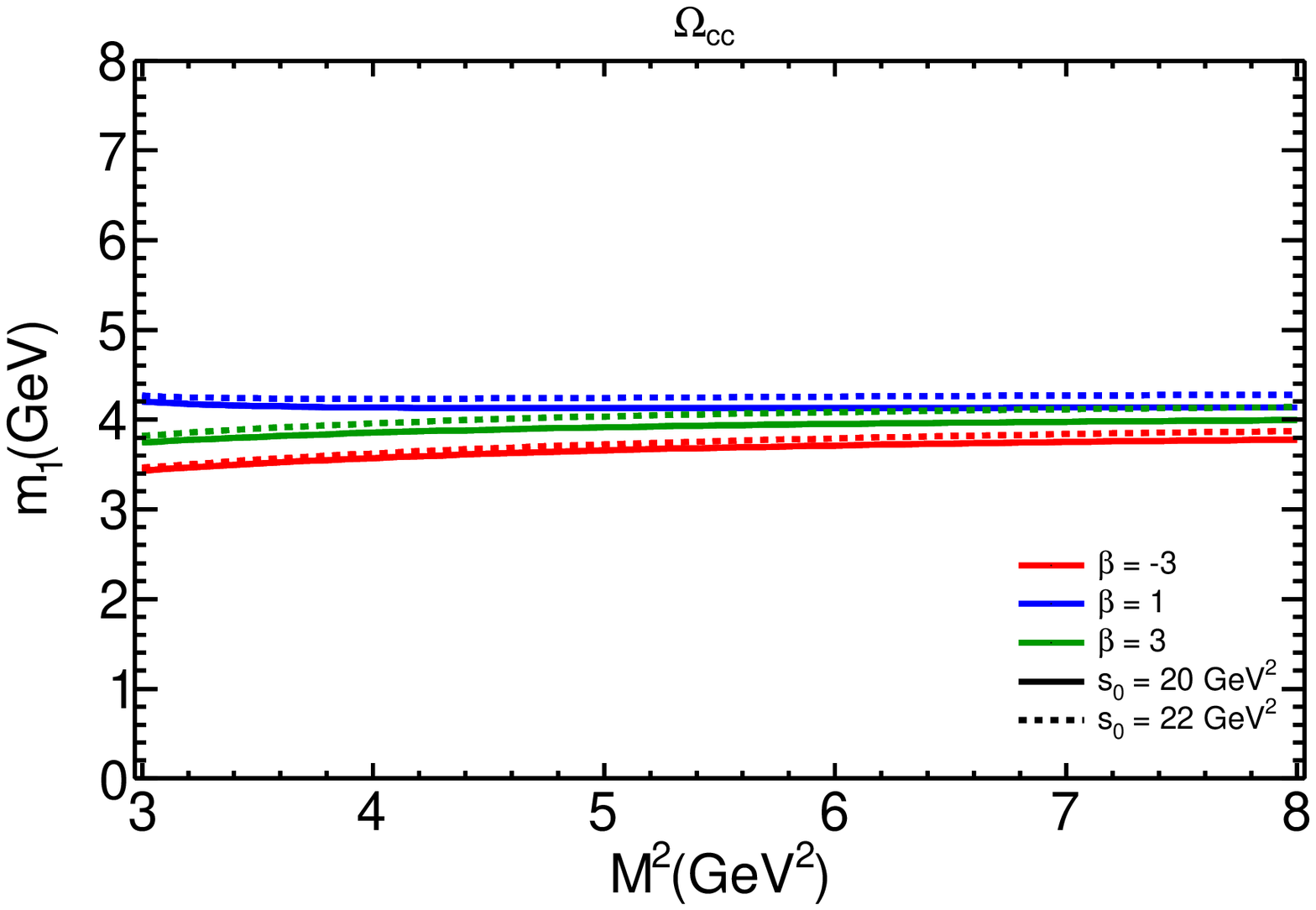}
   \vspace{2em}
  % \caption{~~}
\label{fig:1b}
\end{subfigure}
\begin{subfigure}{\linewidth}
%  \hspace{0.4em}
  \centering
    \includegraphics[width=0.60\linewidth]{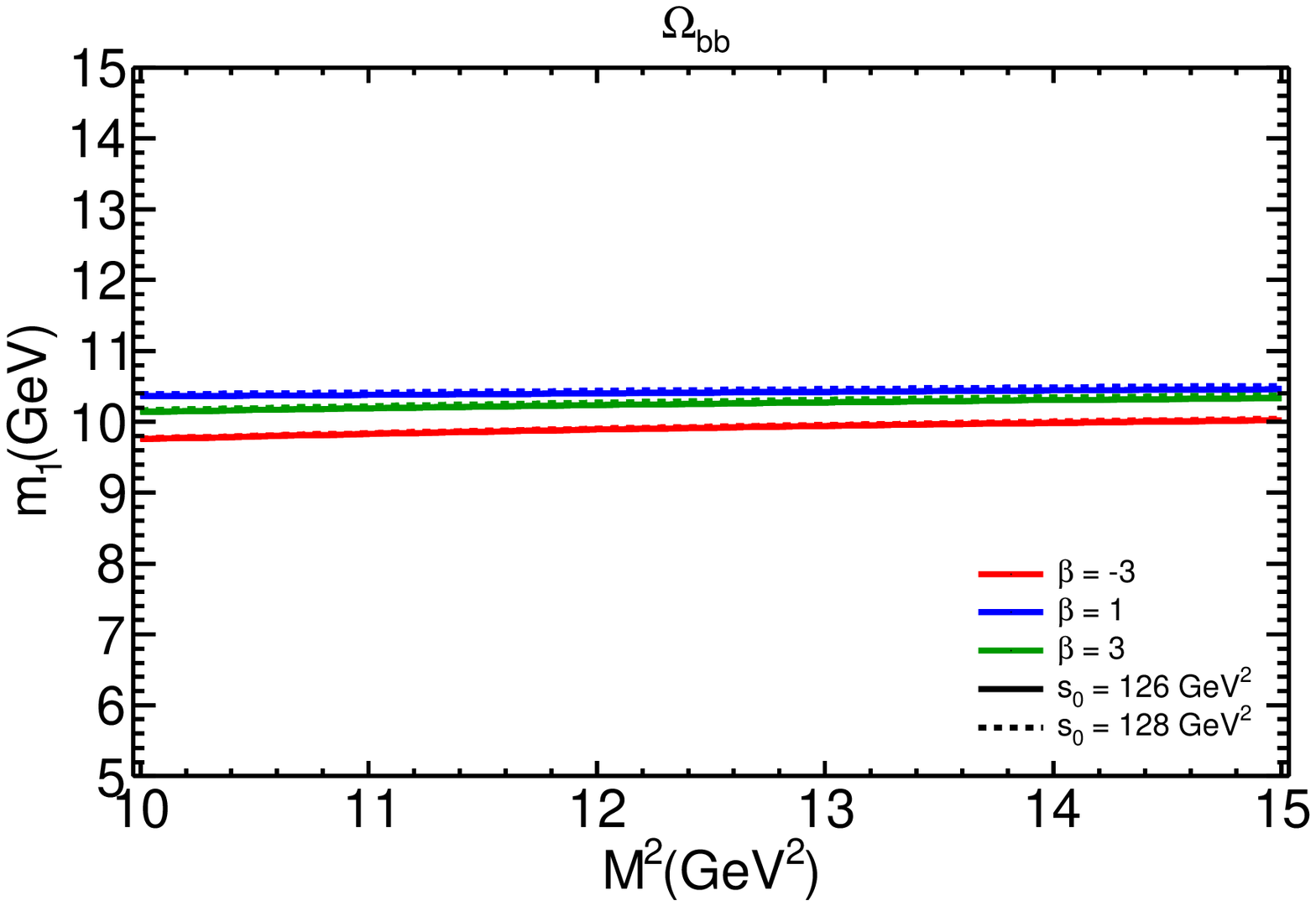}
  % \caption{~~}
\label{fig:1c}
\end{subfigure}
\caption{The dependencies of mass of $2S$ state  $\Xi_{cc}$, $\Omega_{cc}$, and $\Omega_{bb}$ baryons on $M^2$ at fixed values of $s_0$ and $\beta$.}
\label{fig:1}
\end{figure}

\begin{figure}[hbt]
\begin{subfigure}{\linewidth}
\centering
  \includegraphics[width=0.60\linewidth]{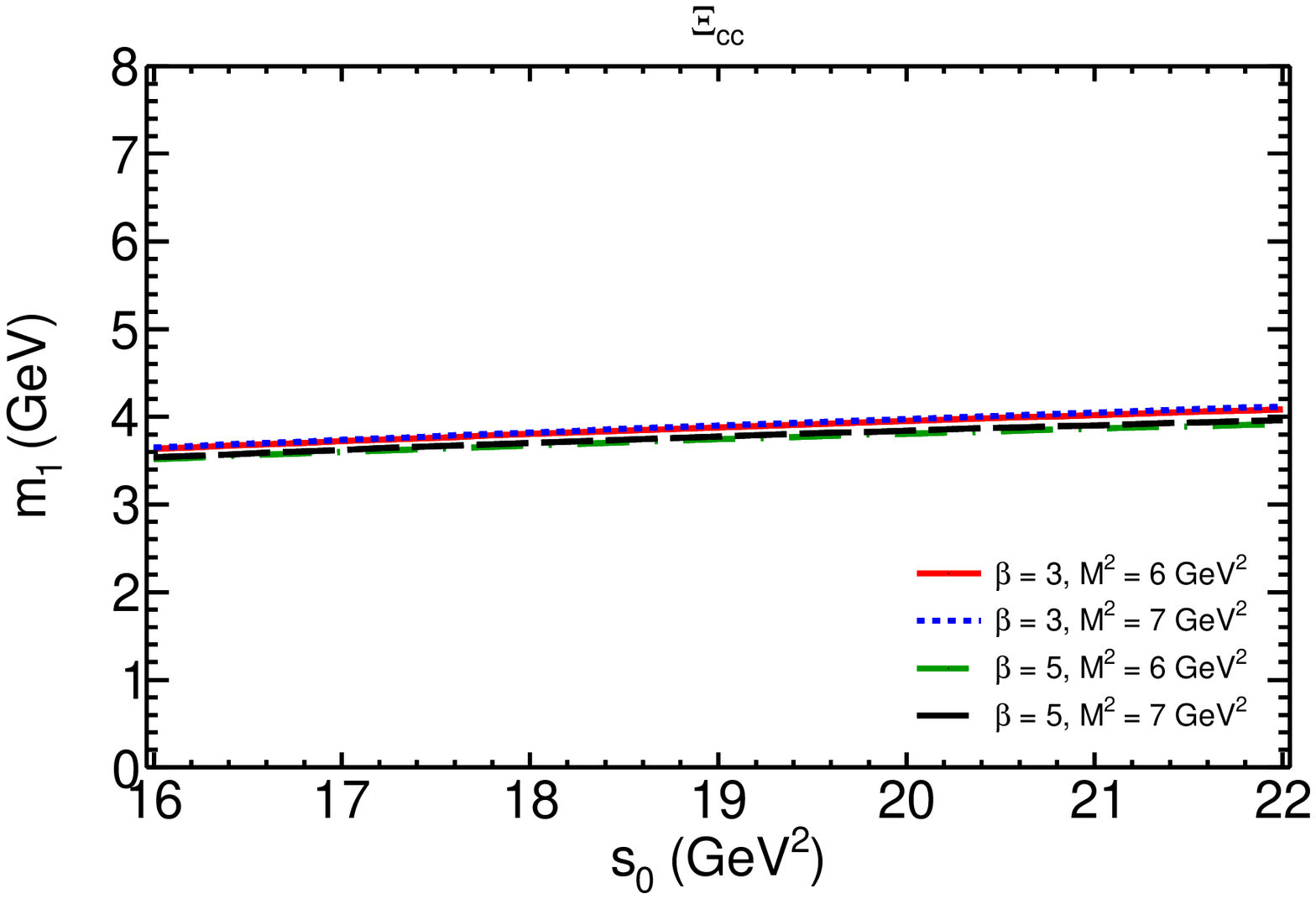}
  % \caption{~~}
\label{fig:2c}
\end{subfigure}%
 \vspace{2em}
\begin{subfigure}{\linewidth}
  \centering
  \includegraphics[width=0.60\linewidth]{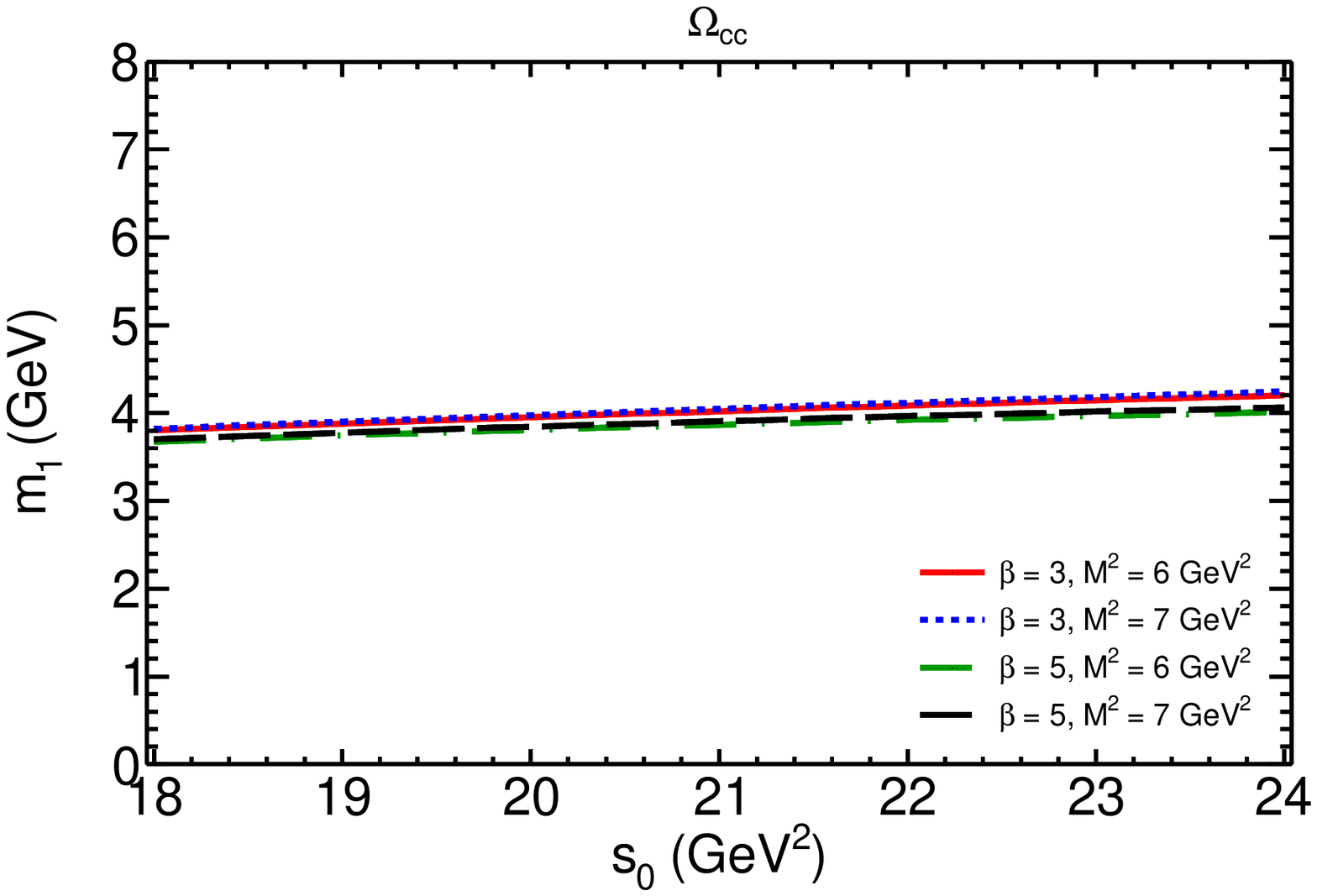}
   \vspace{2em}
  % \caption{~~}
\label{fig:2b}
\end{subfigure}
\begin{subfigure}{\linewidth}
  \hspace{0.4em}
  \centering
    \includegraphics[width=0.60\linewidth]{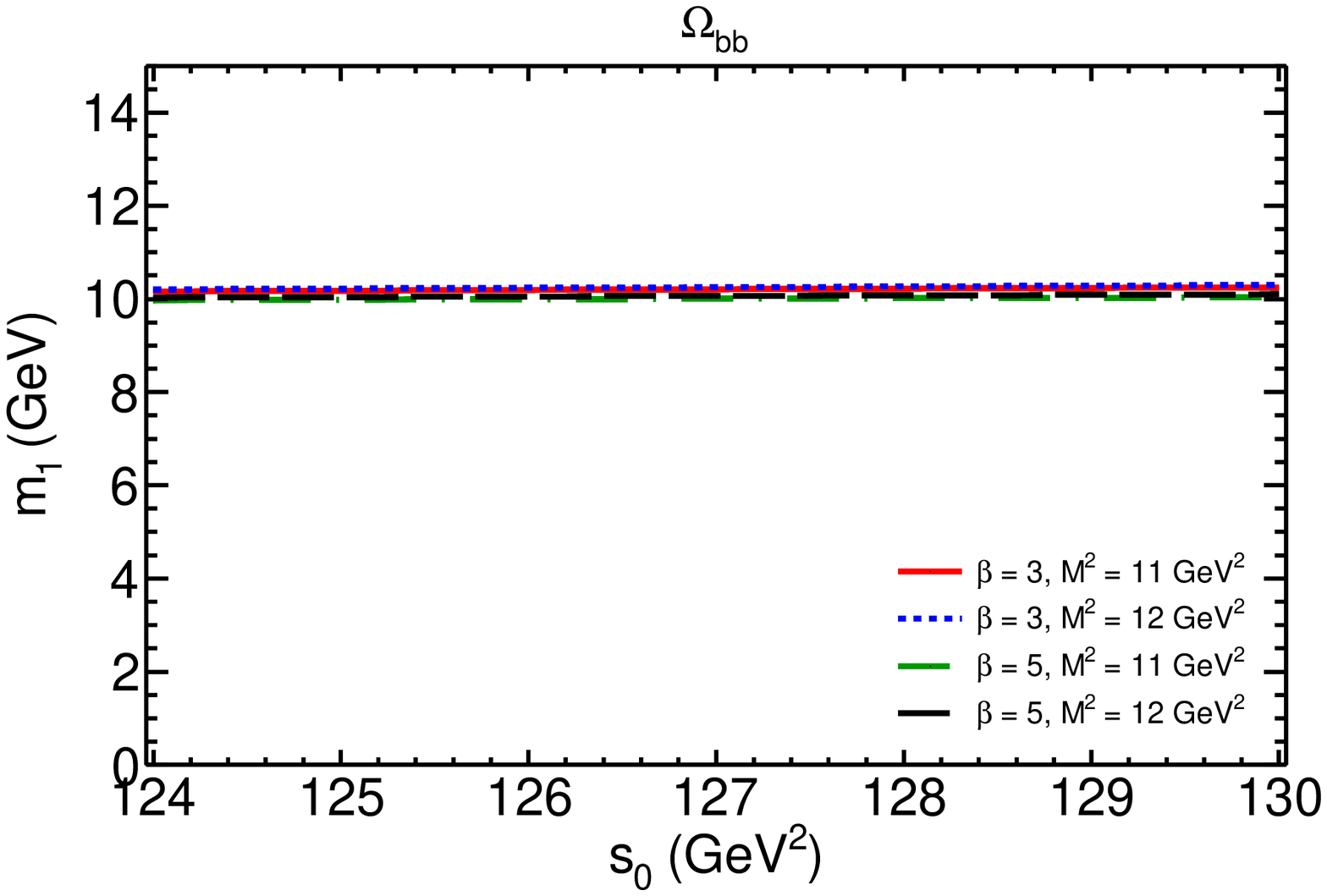}
  % \caption{~~}
\label{fig:2a}
\end{subfigure}
\caption{The dependencies of mass of $2S$ state  $\Xi_{cc}$, $\Omega_{cc}$, and $\Omega_{bb}$ baryons on $s_0$ at fixed values of $M^2$ and $\beta$.}
\label{fig:2}
\end{figure}

\begin{figure}[hbt]
\begin{subfigure}{\linewidth}
\centering
  \includegraphics[width=0.60\linewidth]{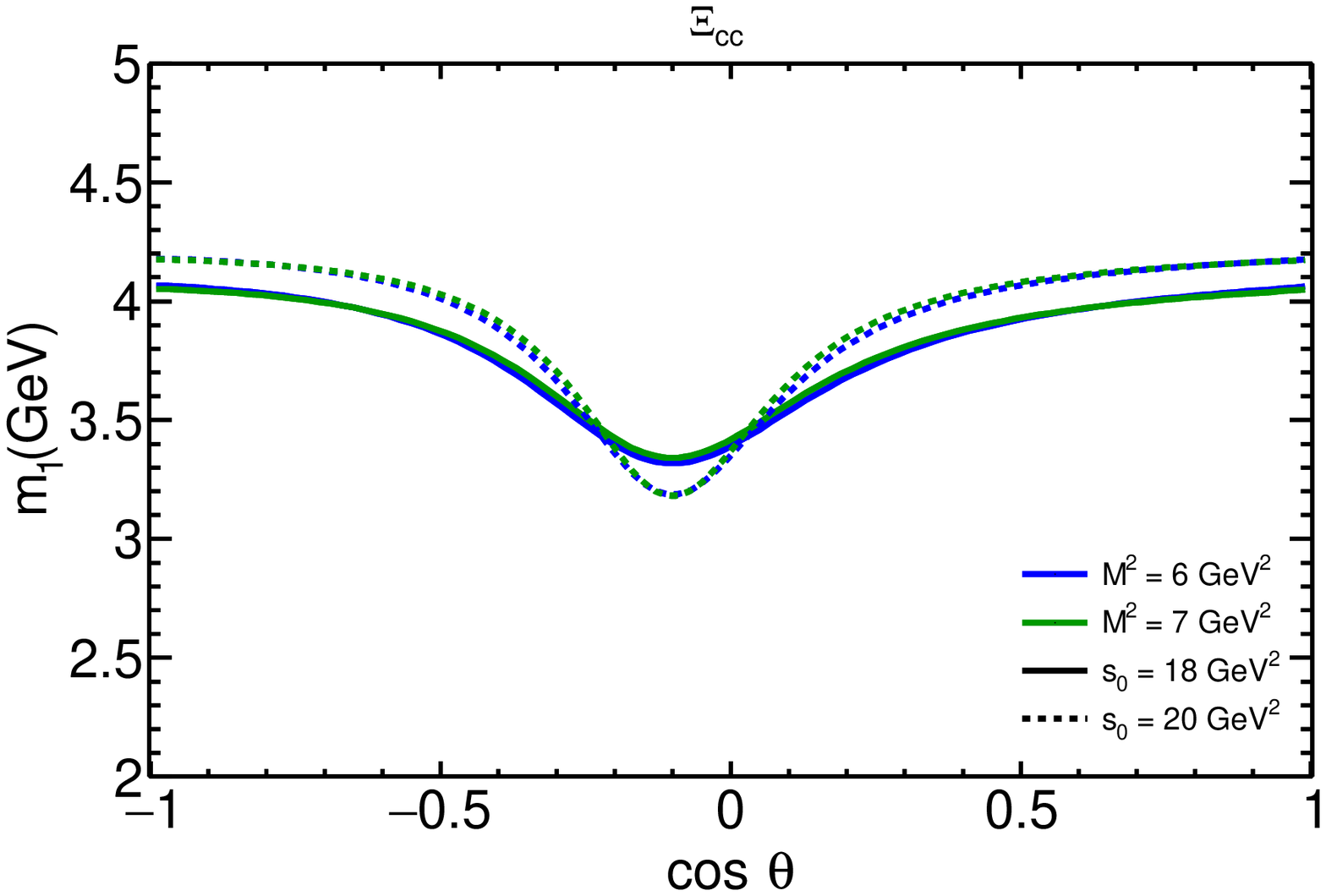}
  % \caption{~~}
\label{fig:4c}
\end{subfigure}%
\vspace{2em}
\begin{subfigure}{\linewidth}
  \centering
  \includegraphics[width=0.60\linewidth]{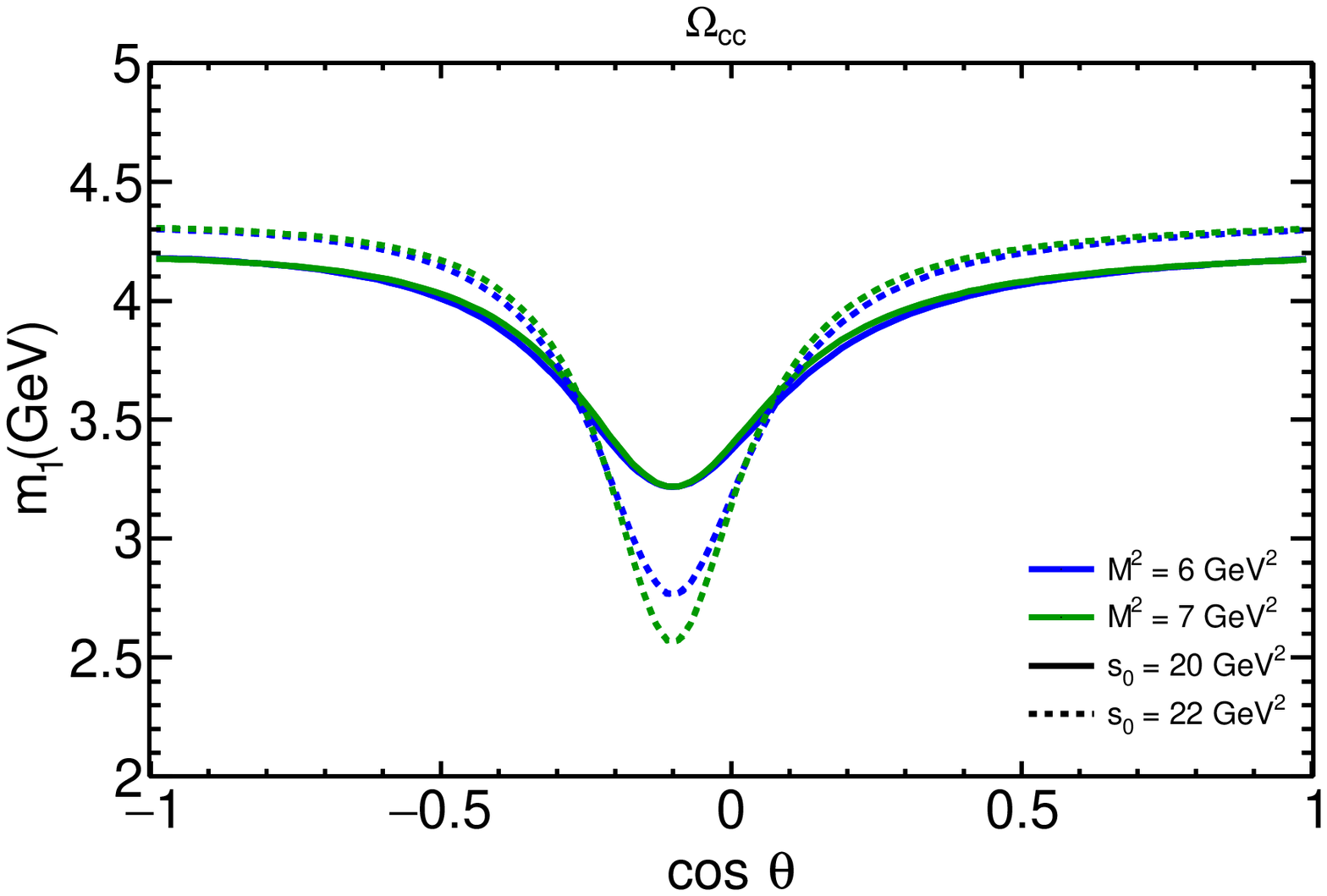}
  \vspace{2em}
  % \caption{~~}
\label{fig:4a}
\end{subfigure}
\begin{subfigure}{\linewidth}
  % \hspace{0.4em}
  \centering
    \includegraphics[width=0.60\linewidth]{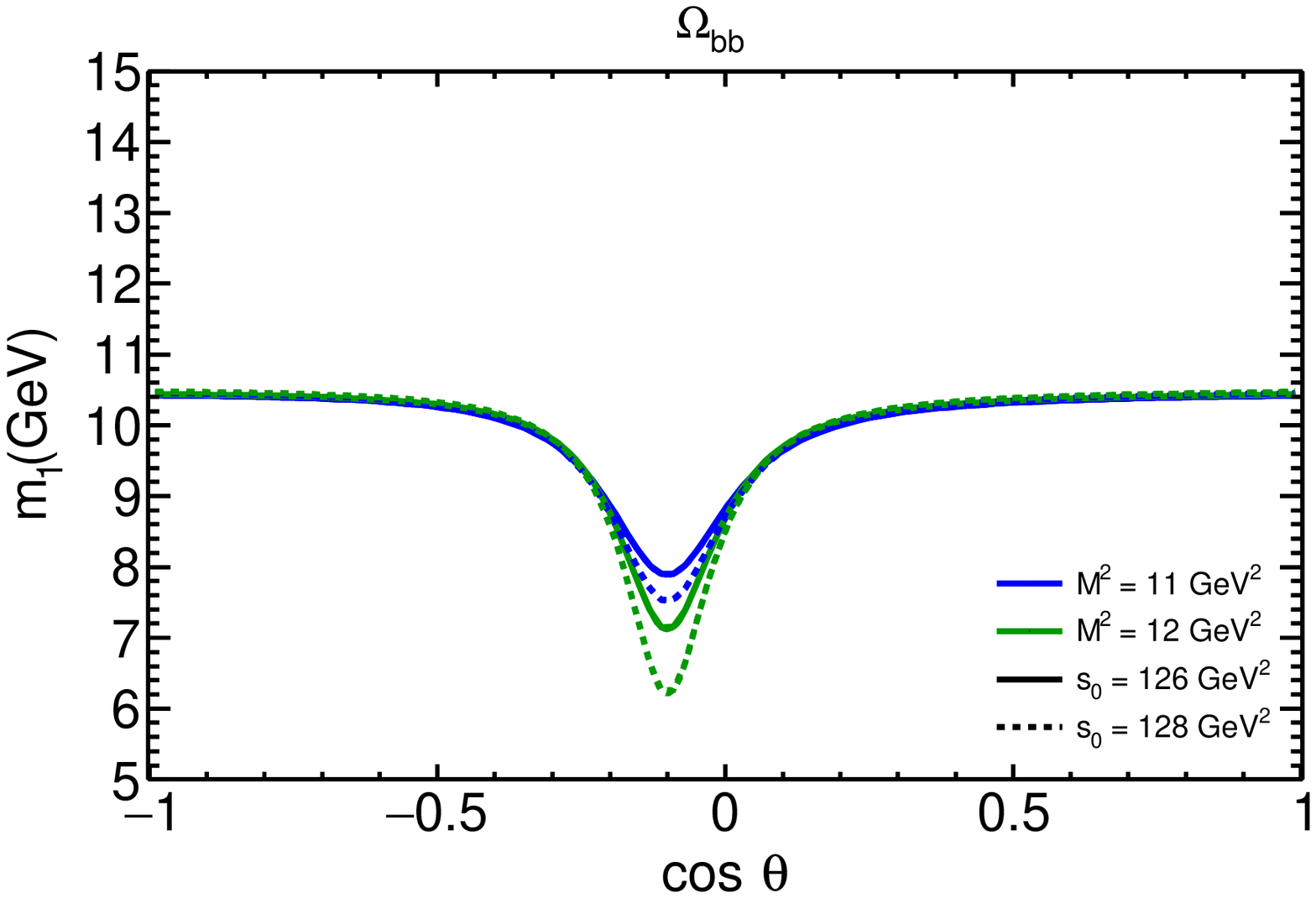}
  % \caption{~~}
\label{fig:4b}
\end{subfigure}
\caption{The dependencies of mass of $2S$ state $\Xi_{cc}$, $\Omega_{cc}$, and $\Omega_{bb}$ baryons on $\cos{\theta}$ at fixed values of $M^2$ and $s_0$.}
\label{fig:4}
\end{figure}

\begin{figure}[hbt]
\begin{subfigure}{\linewidth}
\centering
  \includegraphics[width=0.60\linewidth]{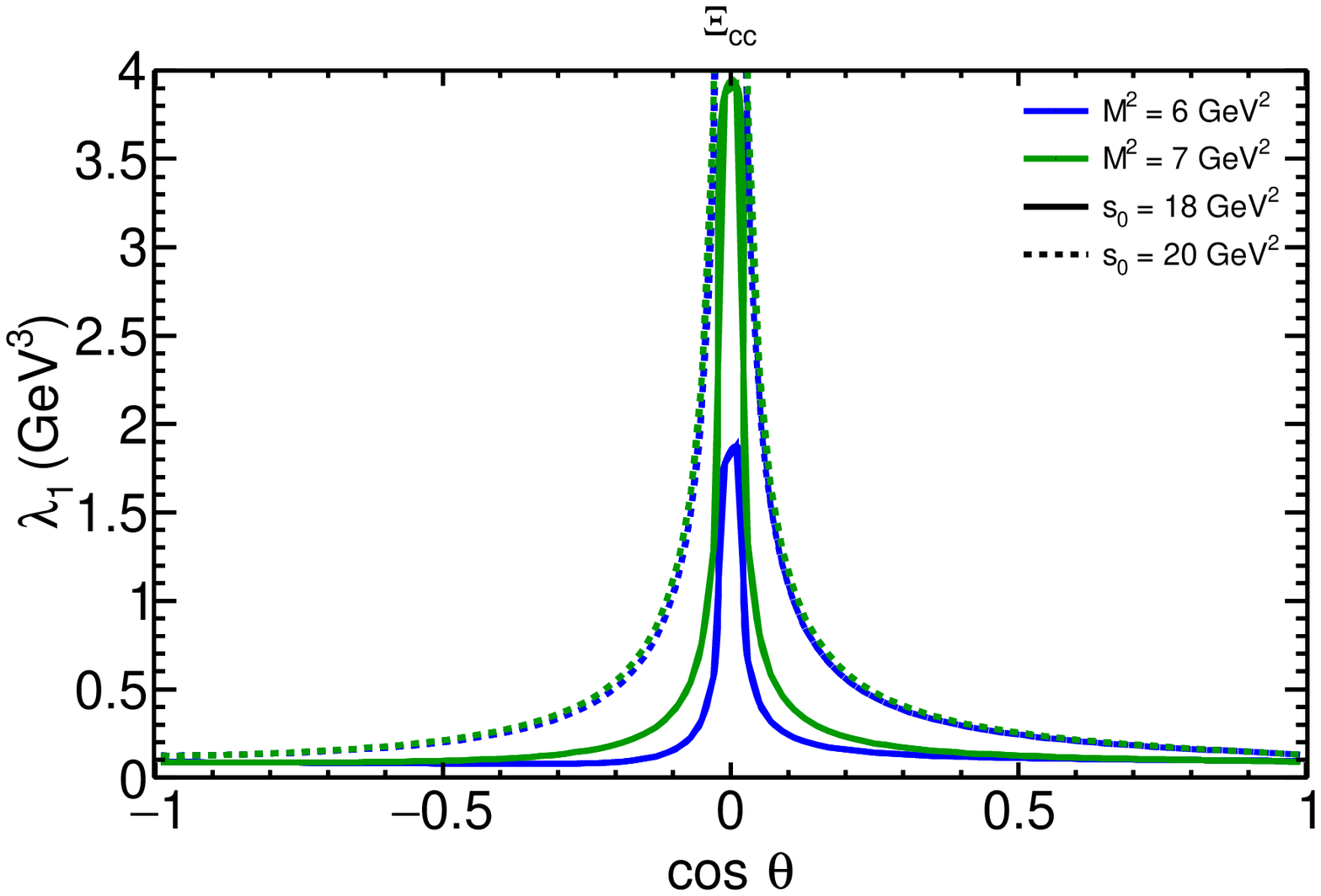}
  % \caption{~~}
\label{fig:3c}
\end{subfigure}%
 \vspace{2em}
\begin{subfigure}{\linewidth}
  \centering
  \includegraphics[width=0.60\linewidth]{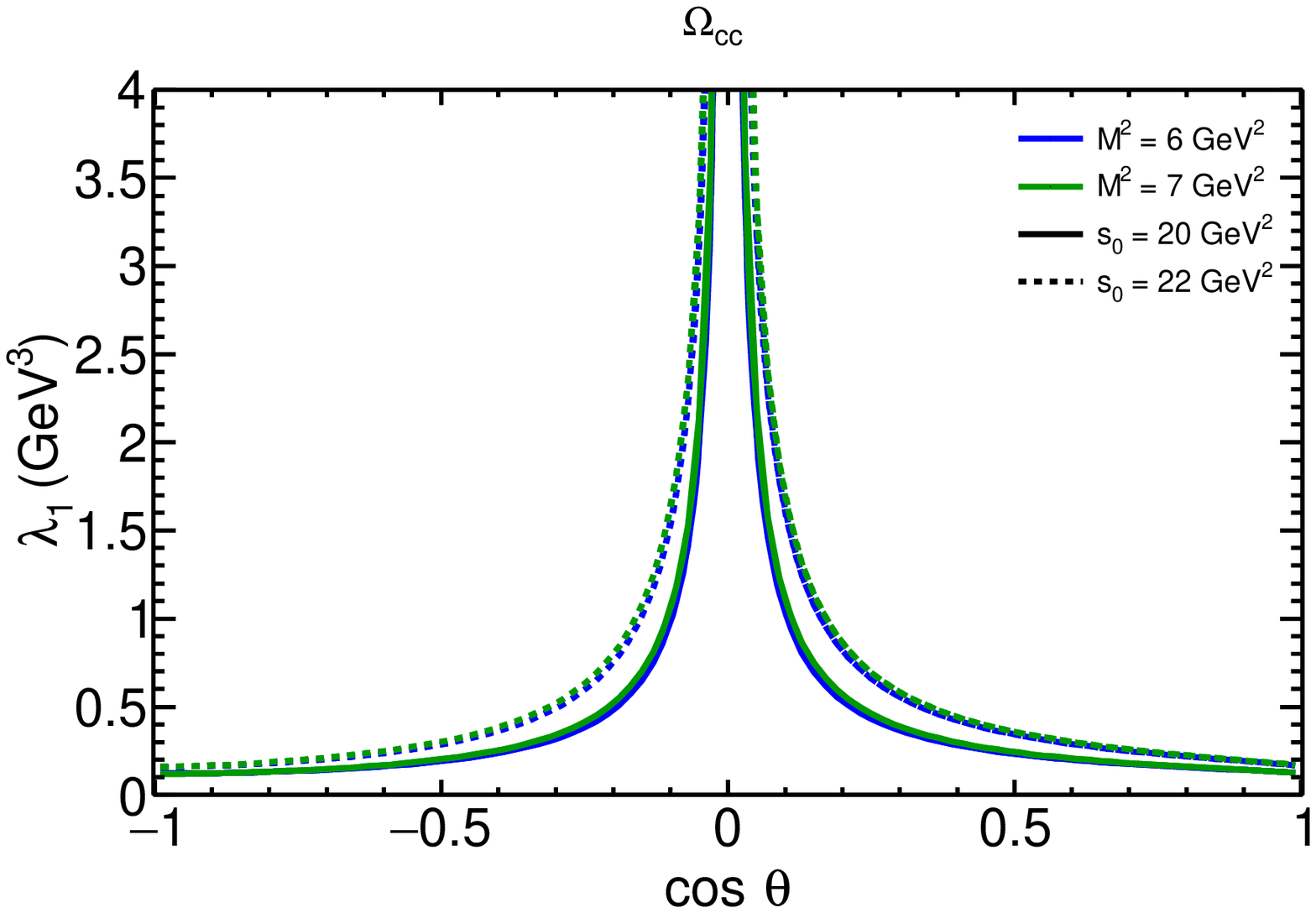}
   \vspace{2em}
  % \caption{~~}
\label{fig:3b}
\end{subfigure}
\begin{subfigure}{\linewidth}
  % \hspace{0.4em}
  \centering
    \includegraphics[width=0.60\linewidth]{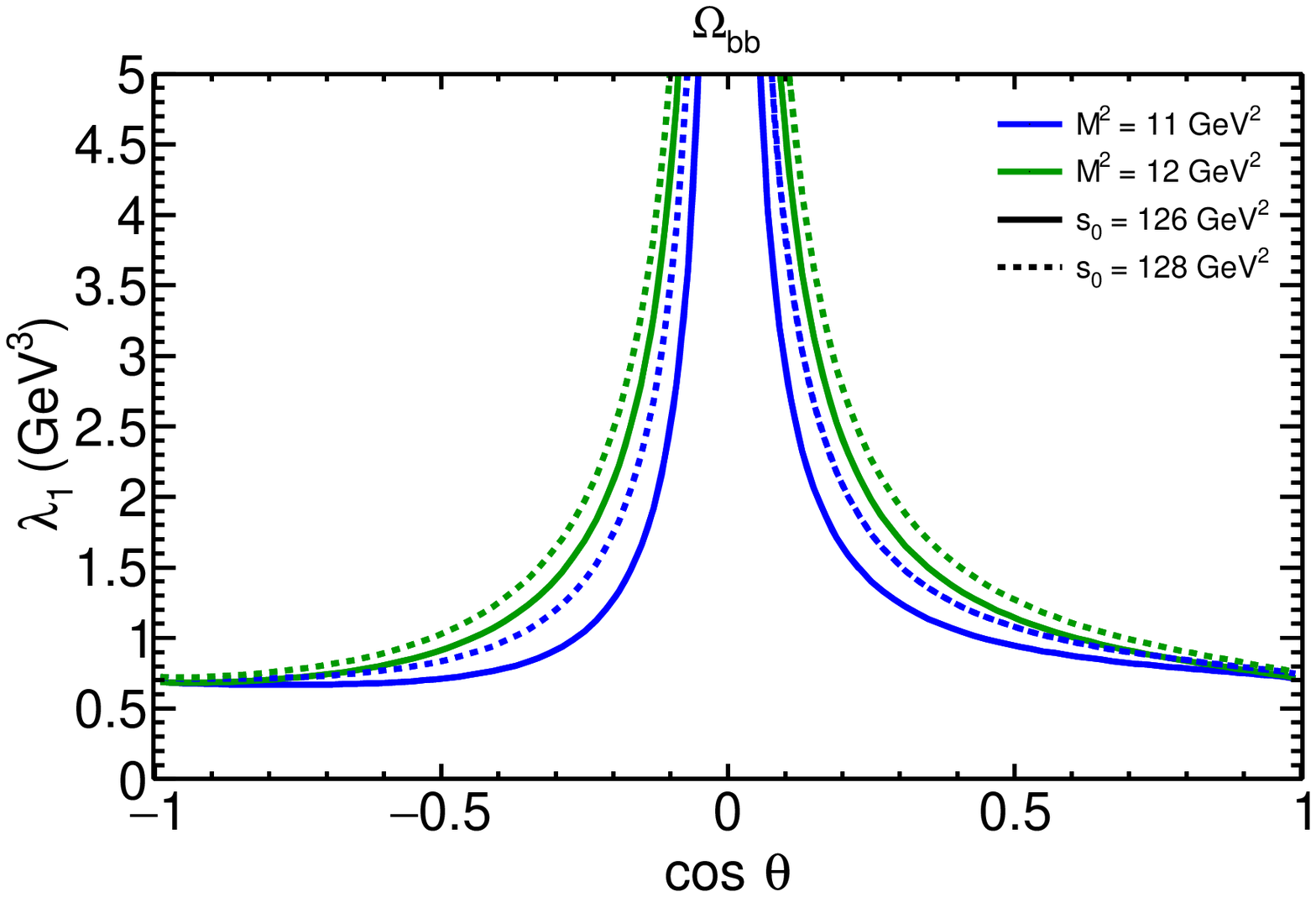}
  % \caption{~~}
\label{fig:3a}
\end{subfigure}
\caption{The dependencies of residue of $2S$ state $\Xi_{cc}$, $\Omega_{cc}$, and $\Omega_{bb}$ baryons on $\cos{\theta}$ at fixed values of $M^2$ and $s_0$.}
\label{fig:3}
\end{figure}

As an example, in Fig.\ref{fig:1}, we present the dependencies of the mass of $2S$ state $\Omega_{cc}$, $\Xi_{cc}$, and $\Omega_{bb}$  baryons on $M^2$ at fixed values of $\beta$ and $s_0$. From this figure, we see that the mass of these baryons  exhibits very good stability to the variation of $M^2$ for its working region. In Fig.\ref{fig:2}, the dependencies of the mass of the aforementioned baryons on $s_0$ at various fixed values of $M^2$ and $\beta$ are depicted. From this figure, it follows that the masses of these baryons are practically unchanged with the variation of $s_0$. Having the working regions of $s_0$ and $M^2$, our final attempt is to find the working region for $\beta = \tan{\theta}$. For this aim in Fig.\ref{fig:4}, we present the dependencies of the mass of $m_{\Xi_{cc}}$, $m_{\Omega_{cc}}$, and $m_{\Omega_{bb}}$ baryons on $\cos{\theta}$ at fixed values of $M^2$ and $s_0$. The results show that the masses of these baryons are practically independent of $\cos{\theta}$ if it varies in the domain  $-1 \leq \cos{\theta} \leq -0.5$ and $0.5 \leq \cos{\theta} \leq 1$. In Fig.\ref{fig:3}, we present the dependence of the residue of $2S$
state $\Omega_{cc}$, $\Xi_{cc}$, and $\Omega_{bb}$ baryons on $\cos{\theta}$ at fixed values of $M^2$ and $s_0$. From this figure,  it is obtained that $\lambda_{1}$ is practically independent of $\cos{\theta}$ when $\cos{\theta}$ varies in the domain $-1 \leq \cos{\theta} \leq -0.5$ and $0.5 \leq \cos{\theta} \leq 1$. In result, we found the common working region for $\cos{\theta}$ in mass and residue predictions of considered baryons. 

Performing similar analysis for the mass and residues for $2S$ and $1P$ excitations of doubly heavy $\Xi$ and $\Omega$ baryons, we get results presented in Tables \ref{tab:2},\ref{tab:3}, and \ref{tab:4}, respectively. For completeness, in this table, we also present the predictions for the mass of doubly heavy baryons obtained from other theoretical approaches. 

\section{Conclusion}
\label{sec:conclusion}
In conclusion, we estimate the masses and residues of the first radial and orbital excitations of doubly heavy baryons. We compare our results with the predictions of other theoretical approaches. Our predictions on mass are in good agreement with other approaches. The obtained results on spectroscopic parameters of excited states doubly heavy
baryons will be useful looking for these states in future experiments.

\section{Acknowledgments}
This study is partially supported under the project METU-GAP-105-2018-2787.
\bibliographystyle{apsrev4-1}
\bibliography{all_qcd}

\newpage
%%%%%%%%%%%%%%%%%%%%%%%%%%%%%%%%%%%%%%%%%%%%%%%%%%%%%%%%%%%%%%%%%%%%%

%%%%%%%%%%%%%%%%%%%%%%%%%%%%%%%%%%%%%%%%%%%%%%%%%%%%%%%%%%%%%%%%%%%%%%%%%%%%%%%%%%%%%%

% %%%%%%%%%%%%%%%%%%%%%%%%%%%%%%%%%%%%%%%%%%%%%%%%%%%%%%%%%%%%%%%%%%%%%%%%%%%%%%%%%%%%%%%%%%%%%%%%%%

% %%%%%%%%%%%%%%%%%%%%%%%%%%%%%%%%%%%%%%%%%%%%%%%%%%%%%%%%%%%%%%%%%%%%%%%%%%%%%%%%%

\end{document}